\documentclass[9pt,twocolumn,twoside]{pnas-report}

\templatetype{pnasresearcharticle}

\usepackage{lipsum}

\title{Node package manager's dependency network robustness}

\author[a,1]{Andrej Hafner}
\author[a]{Anže Mur}
\author[a]{Jaka Bernard}

\affil[a]{University of Ljubljana, Faculty of Computer and Information Science, Ve\v{c}na pot 113, SI-1000 Ljubljana, Slovenia}

\leadauthor{Hafner}

\authordeclaration{All authors contributed equally to this work.}
\correspondingauthor{\textsuperscript{1}To whom correspondence should be addressed. E-mail: ah4910@student.uni-lj.si.}

\begin{abstract}
The robustness of npm dependency network is a crucial property, since many projects and web applications heavily rely on the functionalities of packages, especially popular ones that have many dependant packages. In the past, there have been instances where the removal or update of certain npm packages has caused widespread chaos and web-page downtime on the internet. Our goal is to track the network's resilience to such occurrences through time and figure out whether the state of the network is trending towards a more robust structure. We show that the network is not robust to targeted attacks, since a security risk in a few crucial nodes affects a large part of the network. Because such packages are often backed up by serious communities with high standards, the issue is not alarming and is a consequence of power law distribution of the network. The current trend in average number of dependencies and effect of important nodes on the rest of the network is decreasing, which further improves the resilience and sets a positive path in development. Furthermore, we show that communities form around the most important packages, although they do not conform well to the common community definition using modularity. We also provide guidelines for package development that increases the robustness of the network and reduces the possibility of introducing security risks.

\end{abstract}

\dates{The manuscript was compiled on \today}
\doi{\href{https://ucilnica.fri.uni-lj.si/course/view.php?id=183}{Introduction to Network Analysis} 2020/21}

\begin{document}

\maketitle
\thispagestyle{firststyle}
\ifthenelse{\boolean{shortarticle}}{\ifthenelse{\boolean{singlecolumn}}{\abscontentformatted}{\abscontent}}{}

\dropcap{W}ith the increasing popularity of the JavaScript programming language and the growing community of developers that utilize its power on both the front end and the back end, it is important to know the ins and outs of its possible vulnerabilities. Node package manager (npm) is a package manager for the JavaScript programming language and its runtime environment Node.js and represents a big part of the JavaScript language ecosystem and with that a possible vulnerability. The publicly available packages are open-sourced and developed by the community, so the manager itself doesn't provide any vetting process to identify the quality of the published packages. This can lead to many low-quality packages with breaking changes or bugs and can expose the software to potential security holes and malicious attacks.

In 2016 the removal of one small package with only 11 lines of code from the npm registry triggered the collapse of a large number of packages including very important packages such as Babel. The removal forced npm to change its policy about removing packages. The year 2018 was not a good year for the npm corporation because of three major malicious attacks. At the start of the year in February an issue caused users to permanently break their operating systems by changing the ownership of their system files. In July a certain package was stealing user's credentials and uploading them to the attacker and in November a package was used to steal bitcoins from users. In 2020 another incident occurred where a fault in a small package resulted in an outage in serverless web pages that relied on its functionality in order to operate properly.

We focus on analyzing three aspects of npm dependency network: robustness, its evolution through time and community formation around core packages. Our goal is to see if the current state of the network is prone to cascading failures, in case a security issue occurs in one of the more important packages. Next, we want to analyze how the network evolved through time and whether the current trend is showing improvements in terms of robustness of the network. Finally, we analyze if communities form around core npm packages, on which a large portion of other packages depend.
% With insight into many past incidents, our project proposal introduces three different hypotheses that can shed some light on the robustness of the npm dependency network and provide some insight on the direction in which the development of new packages is going. The results of these hypotheses if proven will serve as a warning to the developers to carefully choose which packages should they include in their software but also as a guideline on how to properly develop new packages with zero or with a low number of dependencies.

% HIPOTEZE OMENI

\begin{figure}[t]\centering%
	\includegraphics[width=0.8\linewidth]{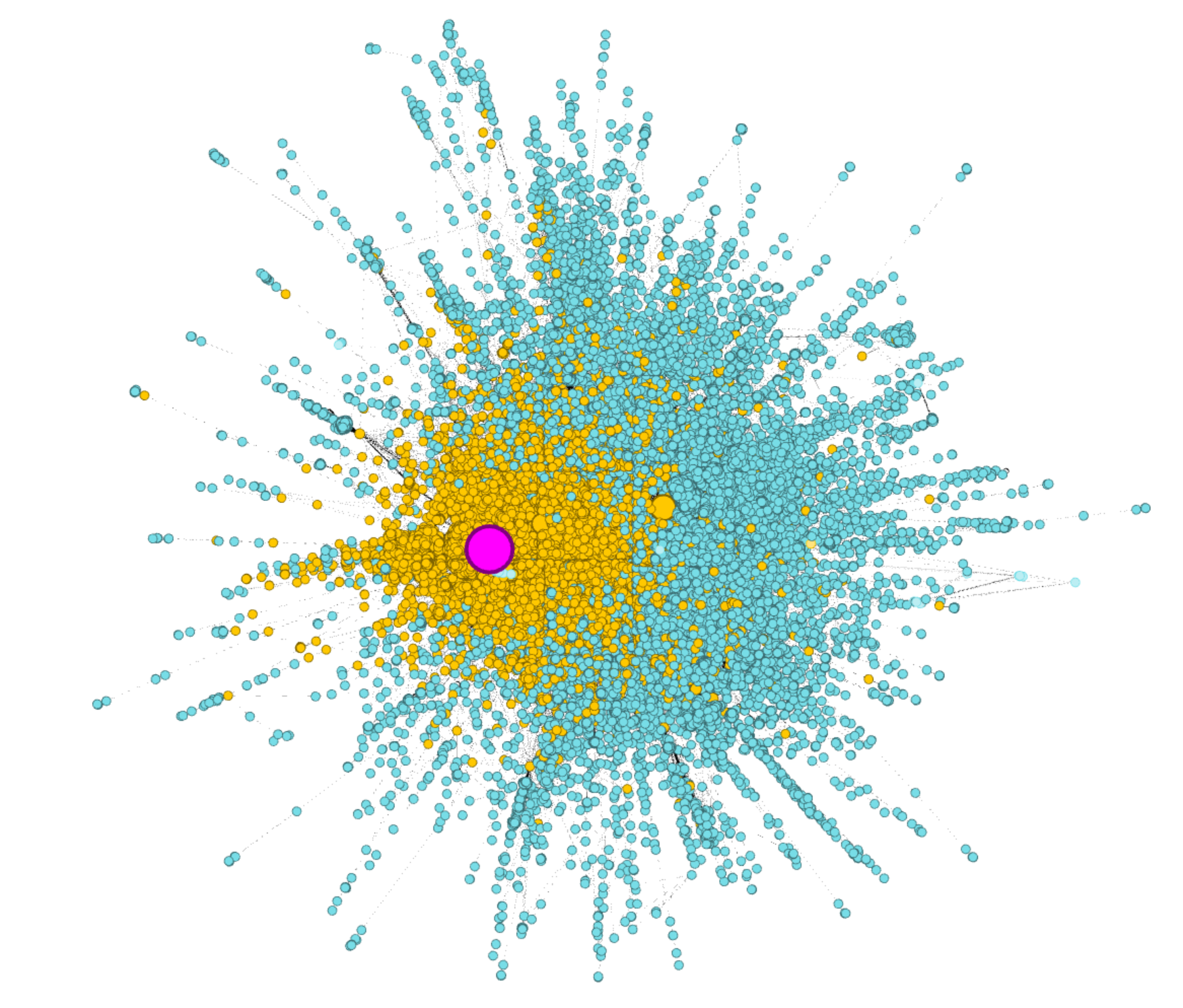}
	\caption{\textbf{A part of vue package neighbourhood contained in a community detected by Louvain community detection algorithm.} All nodes are part of the detected community, node Magenta is the Vue package and orange are nodes in vue's 2-step neighborhood. Vue package is the third most important package in the whole npm dependency network by the PageRank score.}
	\label{fig:vue_comm_neigh}
\end{figure}

\section*{Related work}
Evaluation of network robustness was explored on many different networks with different purposes behind the research. In one paper \cite{callaway2000network}, the authors perform random and targeted node removal experiments on randomly generated networks with varying node degrees and other relevant parameters. Authors of \cite{wang2011modeling} perform similar simulations to \cite{callaway2000network},  but with a different network - they used a complex software network for which they showed that it is more robust than random and scale-free networks to each type of simulated attack. A large number of robustness measures are tested on various different graphs in \cite{ellens2013graph, liucomparative}, with some of the authors further testing the robustness measures after optimizing the graphs. Similarly, the authors of \cite{beygelzimer2005improving} looked at improving robustness through modest linkage alterations, since more invasive network changes are normally not feasible in real-life networks. In \cite{fortuna2011evolution} the authors analyze the package dependency network of the Debian GNU/Linux operating system. They have shown that the modularity of the network of dependencies increased over time, due to aspiration for minimizing project costs. Although modularity did not decrease the number of incompatibilities within the modules, it does increase the fraction of random packages working properly in a random computer. As shown in \cite{nguyen2010studying}, only a small subset of dependency network measures on the level of a programming language predict the post-release failures of software. Authors of \cite{decan2018impact} analyzed the impact of security vulnerabilities in the npm package dependency network. Their work focused on a concrete list of 400 security reports and how they affect the rest of the network. Authors of \cite{guillaume2004comparison} explore and compare the resilience of the scale-free network and randomly generated networks and propose two different attack strategies.

\section*{Results}

%\begin{table}[t]\centering%
%	\caption{Table describing data or methods.}
%	\begin{tabular}{lccccc}\toprule
%	    & $n$ & $m$ & $\langle k\rangle$ & $\langle C\rangle$ & $\langle d\rangle$ \\\midrule
%	    Fine network & $438\,920$ & $9\,742\,733$ & $44.4$ & $0.37$ & $6.19$ \\
%	    Random graph & $438\,920$ & $9\,781\,609$ & $44.6$ & $0.00$ & $4.92$ \\\bottomrule
%	\end{tabular}
%	\label{tbl:example}
%\end{table}

\subsection*{Effects of node failures on network robustness}
\begin{figure}[t]\centering%
	\includegraphics[width=\linewidth]{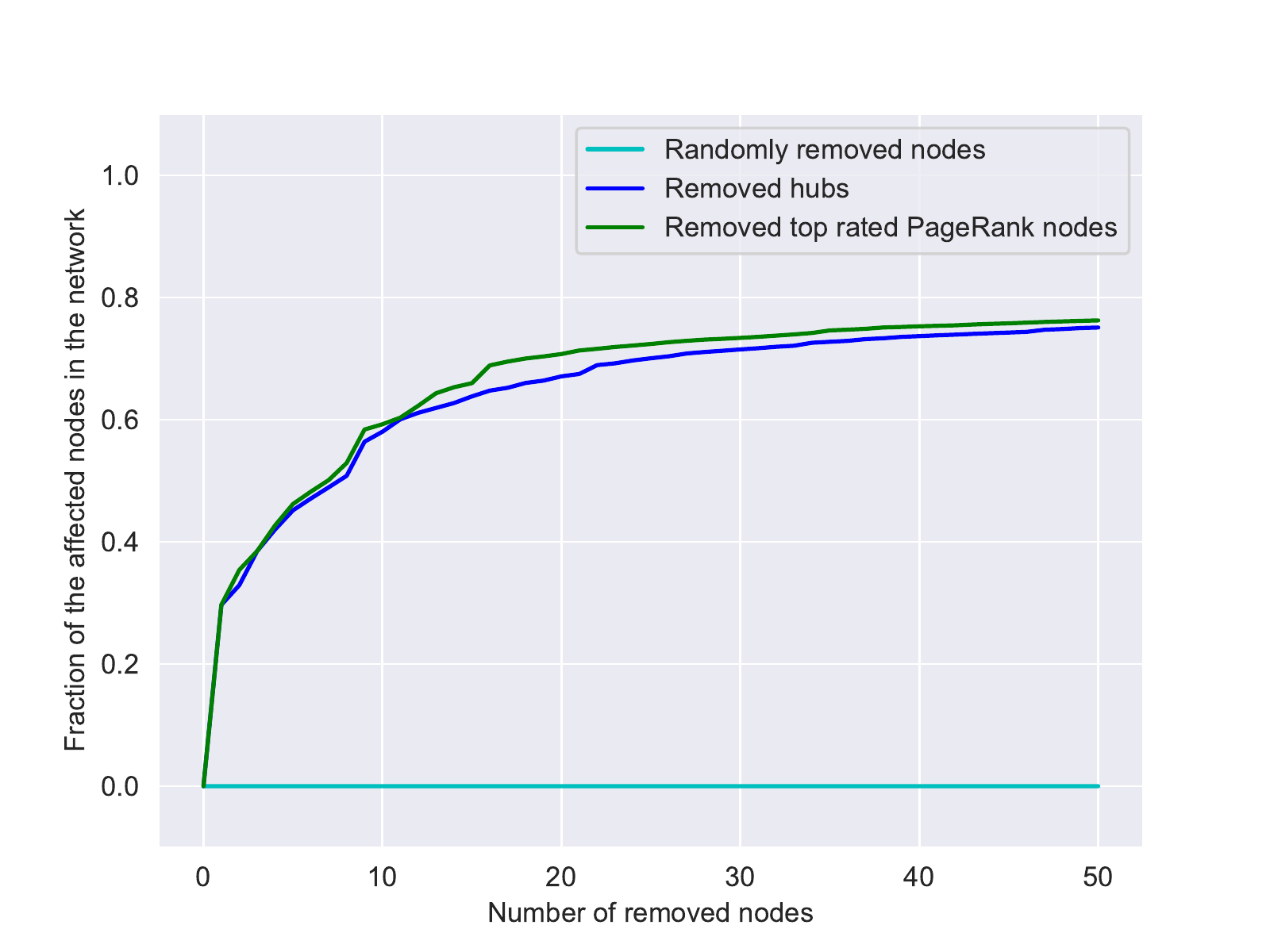}
	\caption{\textbf{Robustness of the npm network.} In the figure we can observe the robustness of the network during the simulated failure of 50 hub nodes, random nodes, and the highest PageRank evaluated nodes. We can see that the network is not at all robust to the failure of hub and PageRank evaluated nodes since with the removal of only one node we observe a network outage of 29.7\%. After the removal of 30 hubs, the network outage starts to slowly stabilize around 90\% of affected nodes.}
	\label{fig:robustness}
\end{figure}

The network robustness was tested on three different cases: failure of random nodes, hub nodes, and important nodes (according to the PageRank \cite{page1999pagerank} score). We treat a node failure as a security vulnerability or an error in the package that prevents it from working correctly. When a node fails the failure spreads across the nodes that are dependant on it and the failure spreads to the deeper levels of dependencies until it reaches all of the connected dependant nodes. From figure (\ref{fig:robustness}) we can see that after the removal of only one node we have already lost 29.7\% of our network. We reach a 60\% network outage after the removal of 12 hubs and their corresponding dependants. After the removal of 130 hubs or important nodes, we reach the network outage of 80\% and after this point, the percentage of the affected nodes starts increasing slower until it stabilizes around 90\% of outage after the removal of around 8000 nodes. This means that there are only 10\% of the nodes left in our network and that they are weakly dependant or not dependant on each other anymore since the removal of the node doesn't have a significant impact on the whole network. From the figure (\ref{fig:robustness}) we can see that our network is totally robust to random node failures but when we removed 10\% (78,233) of the nodes we observed a network outage of 69\%.

We also tested the robustness in terms of connectivity on the same set of targeted nodes. The difference is that connectivity is not measured by the fraction of the failure-affected nodes, but by the fraction of nodes in the largest connected component (LCC) and the failure is not propagated to the dependant nodes \cite{guillaume2004comparison}. Our whole network consists of one LCC but after the removal of 10\% of hubs, a fraction of LCC lowers to 3.8\%, and with the removal of 10\% of PageRank evaluated nodes our LCC is almost empty (0,0001\%). After the removal of 20\% of the hubs, our LCC disintegrates since it includes only 0,00001\% of the nodes and our network is not connected anymore. After the removal of 20\% of the hub nodes on the random graph with the same properties as our network, our LCC was still intact. After the removal of 50\% of hubs fraction of the LCC only lowers by 3.9\% and after the removal of the same number of PageRank evaluated nodes the fraction of LCC only lowers by 13.6\%. 

\subsection*{Network robustness evolution}
\begin{figure}[h!]\centering%
	\includegraphics[width=\linewidth]{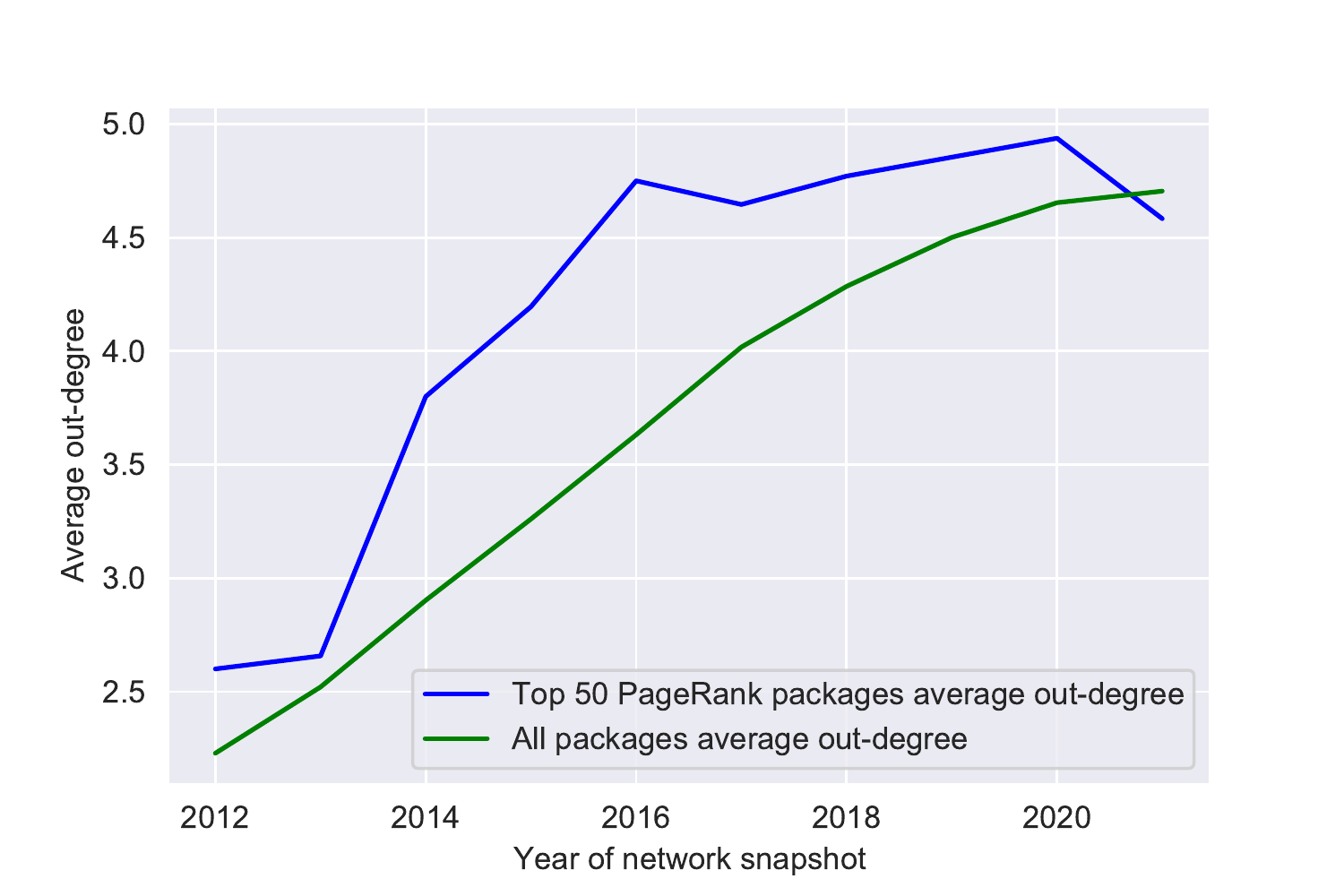}
	\caption{\textbf{Average number of dependencies from 2012 to 2021.} In the figure we show the average out-degree (number of dependencies) of all packages  and the top 50 most important packages by PageRank. Values are show for each of the yearly network snapshots. We can see that the number of dependencies has been increasing throughout the years and is slowing down. Average number of dependencies of the top packages dropped below the average of the whole network for the first time in year 2021.}
	\label{fig:avg_outdegree}
\end{figure}

The evolution of network robustness is first analyzed by comparing the average number of dependencies (out-degree) of all nodes and the 50 most important nodes determined by PageRank. In figure (\ref{fig:avg_outdegree}) we can see the average values for each network snapshot from 2012 to 2021. We can see that in the beginning, the average number of dependencies has been growing steadily for both the whole network and the top 50 packages. In the last few years, the growth started to slow down, especially for most important packages in the network. If we looked directly at the values for the top 50 packages, we would see that most of the packages have 2 or fewer dependencies, with a few having 20 or more (e.g. a famous web server package \textit{express}).

\begin{figure}[h!]\centering%
	\includegraphics[width=\linewidth]{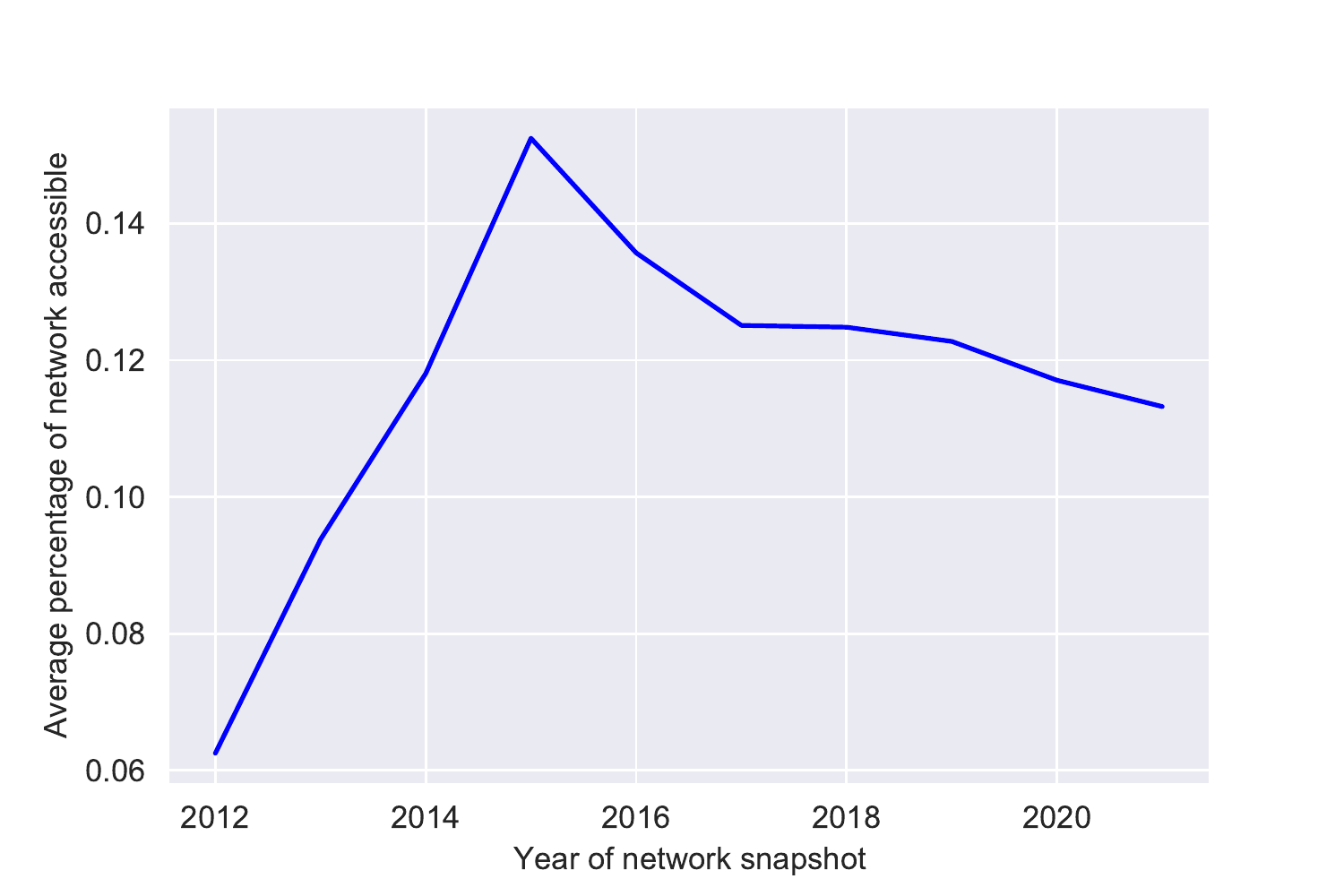}
	\caption{\textbf{Average percentage of the network that depends on the 100 most important packages by PageRank.} In the figure we show what percentage of the network is affected by the most important packages. We can see that the influence of the most important packages was increasing until the year 2015, after which it started gradually dropping.}
	\label{fig:perc_network accessible}
\end{figure}

We also analyze the percentage of the network that depends on the 100 most important packages by PageRank through time. In figure (\ref{fig:perc_network accessible}) we show how the fraction of the affected network changes from year 2021 to year 2021. In the beginning, the percentage was increasing steadily until the year 2015, after which it starts dropping.

\subsection*{Community formation around core packages}
When calculating the intersection ratio between popular package communities and their neighbourhoods, we found that 2-step neighborhoods provided the best ratios between the fraction of the intersection in the community and the fraction in the node neighborhood.

Two examples that stand out are the packages \textit{vue} and \textit{react}, as the intersection covers around half of both the community and neighborhood sub-graphs, as seen in table (\ref{tbl:fractions}) and figure (\ref{fig:vue_comm_neigh}).
Both of these packages, as well as many others among the top of the PageRank ranking also have little to no dependencies or are dependent mostly on their own modules.

Out of the tested 20 top-ranked packages according to PageRank, only three had over 10 dependencies, while most had 3 or less.
Out of the tested packages, $6$ had a community fraction above $40\%$, with the number increasing to $9$ if we lower the bar to $30\%$.
When looking at neighborhood fractions, $9$ had a fraction above $40\%$, with the number increasing to $16$ if we lower the bar to $30\%$.

\begin{table}[h!]\centering%
	\caption{Intersection fractions between communities and popular package neighborhoods.}
	\begin{tabular}{lccc}\toprule
	    & Frac. of community  & Frac. of neighborhood & Dependencies  \\\midrule
	    vue & $49.1\%$  & $67.1\%$ & $0$ \\
	    react & $48.5\%$ & $58.1\%$ & $2$ \\\bottomrule
	\end{tabular}
	\label{tbl:fractions}
\end{table}

%\begin{figure}[h]\centering%
%	\includegraphics[width=0.49\linewidth]{example1}
%	\includegraphics[width=0.49\linewidth]{example2}
%	\caption{Figure showing interesting examples.~\cite{Sub18a}}
%\end{figure}

%\begin{figure}[t]\centering%
%	\includegraphics[width=\linewidth]{distributions}
%	\caption{Figure showing relevant results.~\cite{Sub18a}}
%\end{figure}

%\begin{figure}[b]\centering%
%	\includegraphics[width=0.45\linewidth]{results1}\hskip12pt
%	\includegraphics[width=0.45\linewidth]{results2}
%	\caption{Another figure with results.~\cite{Sub18a}}
%	\label{fig:example}
%\end{figure}

\section*{Discussion}

From the figure (\ref{fig:robustness}) we can see that our network is not robust at all to the failures of hubs and nodes evaluated by PageRank since after the removal of only one node we have already lost almost one-third of our network. Even more, our network experience an outage of 90\% after only 1\% of hubs is targeted by a malicious attack. We observed that the removal of the hubs and very important nodes have almost the same impact on the network with the impact of PageRank evaluated nodes being slightly higher (the average difference between them is 0.004\% based on 78,233 calculations). The reason behind this is that hubs have high in-degrees and are therefore highly ranked by PageRank, as the algorithm gives higher value to nodes that are depended on frequently or are depended on by other high-ranked nodes. With the removal of 10\% of randomly selected nodes and their dependants, we observed a large network outage of 69\%. The reason behind this is that our network is very sensitive to the removal of hub nodes, and until one of these randomly selected nodes is not a hub it looks like our network is highly robust to random node failures. Otherwise, we can observe big spikes in node outages. Because of that, we can't really say that our network is robust to random node failures.

In terms of connectivity robustness, we can say that our network is also not robust at all as its connectivity collapses after the removal of only 10\% of the PageRank evaluated nodes. We used the comparison to the random graph to check that if this behavior of our network isn't random but an actual consequence of our network's structure. This behavior is expected and the reason behind our network's rapid disintegration is that it follows a power-law in-degree distribution. With power-law distribution, the higher the degree, the lower the probability of occurrence of a node with such degree, so the number of nodes with a high degree is low. By targeting these high degree nodes, their numbers are quickly reduced and the network loses a lot of its connectedness and therefore collapses.

Results of network robustness evolution show that the average number of package dependencies has been increasing since the creation of the package manager, and the growth is starting to slow down. This corresponds with the initial trend in web software development, where using packages cut down costs and sped up the development. In the last few years there has been a growing trend of zero-dependency packages, which reduces security risks, although at a cost of longer development. The latter is especially visible in the most important packages, which tend to mostly be famous web frameworks and libraries that are often used with them. Communities that form around them and continue their development seem to lowering the number of dependencies, which can be seen in figure (\ref{fig:avg_outdegree}) with the decline from 2020 to 2021.

When analyzing the percentage of affected network by the most important packages, we see that in the beginning the trend is increasing, then declining after year 2015. We see that on average, the extent of the potential security risks in important package on the network is decreasing, since a continually lesser part of the network depends on them in the past few years. This not only means that important packages are being used less, but also that overall dependence on packages down the line is decreasing. Such trend means a positive change in robustness of the network.

As assumed, popular package neighbourhoods are partially contained in communities detected by the Louvain algorithm.
We can see that large parts of popular package neighborhoods are part of their respective communities, which sometimes intersect between each other.
With community packages depending on core packages or specialized core packages, networks are divided according to the framework they belong to or are made for.
It would make little to no sense for a package dependent on \textit{vue} to also depend on \textit{react}, since they represent competing web development packages.
By following the zero-dependency paradigm, these popular packages reduce the impact of outages, especially if they keep their deployment secure from attacks and unintended errors. Community detection was done using the Louvain algorithm, which might not be the perfect for our network structure. The algorithm optimizes modularity, which means that it searches for clusters that have many connections between the nodes within and little connections between clusters. Such structure does not exactly represent the communities in the npm dependency network, where there are mostly dependency trees, without many connections between the packages that depend on the same package. This is because normally, packages implement a limited functionality or extend the functionality of the package they depend on. Such packages in most cases don't depend on each other, since adding other functionalities is not their goal. Because of this the clusters detected are not optimal, but do partially show communities around packages. Other community detection algorithms were used, but results couldn't be obtained due to memory limitations caused by the rather large network size.

Overall, we show that the current network state is fragile, since security issues introduced into a few of the most important packages effect a large portion of the network. This is probably not alarming, since these packages are often backed up by companies and are not likely to contain serious problem, due to a large community of professionals backing them up. The current trend in the robustness of the network is positive, since the average number of dependencies is stabilizing and even dropping for the most important packages. The latter also affect a continually lesser part of the whole network, which also brings improvements in terms of security. We suggest that developers of the npm packages try to keep the number of dependencies at minimum or mostly use well-known packages which are backed up by a serious community. When specifying dependencies, they should specify exact version and use version locking mechanisms, which prevent automatic updates to newer versions without knowing. Updates should be done manually, checking what has been changed and making sure that security analysis has been passed by updated versions.

\section*{Methods}

\subsection*{Data} The npm dependency network was extracted and parsed from the official npm package registry \cite{npm_reg}. Each node represents a single npm package and an edge between two nodes represents the one-way dependency between them, with a link from package A to package B if A depends on B - the network is directed. In its raw form, it contains 1,741,751 nodes and 3,594,421 edges. Since a large fraction of the nodes are isolated from the main network, we decided to use the largest weakly connected component, which contains 782,332 nodes and 2,572,892 edges. Removed nodes are mostly independent islands and don't change the main network structure, which we analyze. The average degree of our network is 4.567 and the network’s in-degree distribution follows a power-law degree distribution, similar as in \cite{fortuna2011evolution}.

\subsection*{Approaches}
\subsubsection*{Effects of node failures on network robustness}
For the testing of the effects of node failures on network robustness, we use three different types of failure simulations. Random node failures represent random errors while hub and PageRank evaluated node failures represent targeted attacks. In every iteration of our simulated node failure we choose our targeted node - randomly, based on its in-degree or PageRank score. We then create a sub-graph around this node to find all of the nodes that are dependant on it. After that, we remove this sub-graph from our original network and measure the fraction of the affected nodes. With this, we successfully measure the impact of a single node on the whole network. We repeat this procedure until we remove 10\% of targeted nodes. For the connectivity robustness of our network, we use a similar approach the only difference was that we remove our nodes in batches and that the failure of the node isn't spreading to its dependants. After the removal of the batch of targeted nodes, we compute the fraction of the nodes in LCC. The same procedure is then repeated on a randomly generated graph with the same number of nodes and edges as our original graph.

\subsubsection*{Network robustness evolution}
We analyze network robustness evolution by first creating network snapshots. Data obtained from the npm registry contains version history and release times, which enable us to extract the network structure at a certain point in time. Using this approach, we create ten networks ranging from year 2012 to 2021, considering version updates. If there are no new versions for a certain year, we use the last existing version. We then first calculate the average out-degree of all the nodes in the network for each year. This enables us to see how the average number of dependencies change through time. Since the guidelines and trends are going toward packages with as little dependencies as possible, we assume that the value would drop. Furthermore, we calculate the average out-degree of the 50 packages that have the highest PageRank score in each snapshot. This way, we compare how the average number of dependencies (out-degree) of packages in the network compares to the most important packages in the network.
In our second approach, we analyze the effect of security issues of the most important packages on the rest of the network. We again calculate the PageRank values of each network snapshot and calculate the effect of the 100 nodes with the highest PageRank score. The number of top nodes is fixed to 100 in order to always use only the most important packages in the network. If we use a percentage of the current network size, the calculations are not realistic, since the network growth speed isn't the same as the growth of number of important packages. We then calculate the percentage of nodes in the network that depend on these top packages by running depth-first search on each one of them. For each snapshot the percentages are averaged.

\subsubsection*{Community formation around core packages}
For detecting communities, we use the Louvain algorithm.
For this purpose, we consider our graph as an undirected graph, as Louvain does not accept directed graphs.
To see if communities indeed form around popular packages, we take the top 20 packages ranked by PageRank, and extract sub-graphs of distances 1, 2 and 3 around these nodes.
One step represents the packages that depend directly on a core package.
Since the dependence is direct, it would make sense that these packages are in the same community as the core package.
Two steps represent either the core package and its extended dependents or the core package with specialized core packages along with their dependents.
As all packages are still relatively close or directly connected to the core, their inclusion in the same community would also make sense.
Three steps represent the core, the specialized core, and the extended dependent packages of a root package.
Since three steps already cover a large portion of the network, we did not expect there to be a large overlap with the communities, but tested in case our expectations were wrong.
We then look at the communities of the root node, extract that community from the whole graph, then look at the intersecting nodes between the community sub-graph and package neighborhood sub-graph.
From the sizes of these node subsets we compute the fraction that each overlap represents of the community and node neighborhood in question.

%\acknow{The authors would like to\dots}
%\showacknow{}
%\bibliography{bibliography}

\begin{thebibliography}{11}
\providecommand{\natexlab}[1]{#1}
\providecommand{\url}[1]{\texttt{#1}}
\expandafter\ifx\csname urlstyle\endcsname\relax
  \providecommand{\doi}[1]{doi: #1}\else
  \providecommand{\doi}{doi: \begingroup \urlstyle{rm}\Url}\fi

\bibitem[{\v S}ubelj(2018)]{Sub18a}
Lovro {\v S}ubelj.
\newblock Convex skeletons of complex networks.
\newblock \emph{J. R. Soc. Interface}, 15\penalty0 (145):\penalty0 20180422,
  2018.

\bibitem[Kleinberg(2000)]{Kle00}
Jon~M. Kleinberg.
\newblock Navigation in a small world.
\newblock \emph{Nature}, 406\penalty0 (6798):\penalty0 845, 2000.

\bibitem[Bourne(2005)]{Bou05}
Philip~E. Bourne.
\newblock Ten simple rules for getting published.
\newblock \emph{PLoS Comput. Biol.}, 1\penalty0 (5):\penalty0 e57, 2005.

\bibitem[Erren and Bourne(2007)]{EB07}
Thomas~C. Erren and Philip~E. Bourne.
\newblock Ten simple rules for a good poster presentation.
\newblock \emph{PLoS Comput. Biol.}, 3\penalty0 (5):\penalty0 e102, 2007.

\bibitem[Newman(2008)]{New08}
Mark E.~J. Newman.
\newblock The physics of networks.
\newblock \emph{Phys. Today}, 61\penalty0 (11):\penalty0 33--38, 2008.

\bibitem[Fortunato(2010)]{For10}
Santo Fortunato.
\newblock Community detection in graphs.
\newblock \emph{Phys. Rep.}, 486\penalty0 (3-5):\penalty0 75--174, 2010.

\bibitem[Newman(2012)]{New12}
M.~E.~J. Newman.
\newblock Communities, modules and large-scale structure in networks.
\newblock \emph{Nat. Phys.}, 8\penalty0 (1):\penalty0 25--31, 2012.

\bibitem[Fortunato and Hric(2016)]{FH16}
Santo Fortunato and Darko Hric.
\newblock Community detection in networks: {A} user guide.
\newblock \emph{Phys. Rep.}, 659:\penalty0 1--44, 2016.

\bibitem[Peel et~al.(2017)Peel, Larremore, and Clauset]{PLC17}
Leto Peel, Daniel~B. Larremore, and Aaron Clauset.
\newblock The ground truth about metadata and community detection in networks.
\newblock \emph{Sci. Adv.}, 3\penalty0 (5):\penalty0 e1602548, 2017.

\bibitem[Peel et~al.(2018)Peel, Delvenne, and Lambiotte]{PDL18}
Leto Peel, Jean-Charles Delvenne, and Renaud Lambiotte.
\newblock Multiscale mixing patterns in networks.
\newblock \emph{P. Natl. Acad. Sci. USA}, 115\penalty0 (16):\penalty0
  4057--4062, 2018.

\bibitem[Peixoto(2020)]{Pei20}
Tiago~P. Peixoto.
\newblock Bayesian stochastic blockmodeling.
\newblock In Patrick Doreian, Vladimir Batagelj, and Anu{\v s}ka Ferligoj,
  editors, \emph{Advances in {Network} {Clustering} and {Blockmodeling}},
  Computational and {Quantitative} {Social} {Science}, pages 281--324. Wiley,
  New York, 1st edition, 2020.

\end{thebibliography}


\section*{References}
\begin{thebibliography}{11}
\providecommand{\natexlab}[1]{#1}
\providecommand{\url}[1]{\texttt{#1}}
\expandafter\ifx\csname urlstyle\endcsname\relax
  \providecommand{\doi}[1]{doi: #1}\else
  \providecommand{\doi}{doi: \begingroup \urlstyle{rm}\Url}\fi

\bibitem[Callaway et~al.(2000)Callaway, Newman, Strogatz, and
  Watts]{callaway2000network}
Duncan~S Callaway, Mark~EJ Newman, Steven~H Strogatz, and Duncan~J Watts.
\newblock Network robustness and fragility: Percolation on random graphs.
\newblock \emph{Physical review letters}, 85\penalty0 (25):\penalty0 5468,
  2000.

\bibitem[Wang and Liu(2011)]{wang2011modeling}
Jian Wang and Yan-Heng Liu.
\newblock Modeling software faults propagation.
\newblock \emph{EPL (Europhysics Letters)}, 92\penalty0 (6):\penalty0 60009,
  2011.

\bibitem[Ellens and Kooij(2013)]{ellens2013graph}
Wendy Ellens and Robert~E Kooij.
\newblock Graph measures and network robustness.
\newblock \emph{arXiv preprint arXiv:1311.5064}, 2013.

\bibitem[Liu et~al.()Liu, Zhou, Wang, and Liu]{liucomparative}
J~Liu, M~Zhou, S~Wang, and P~Liu.
\newblock A comparative study of network robustness measures. front. comput.
  sci. 11, 568--584 (2017).

\bibitem[Beygelzimer et~al.(2005)Beygelzimer, Grinstein, Linsker, and
  Rish]{beygelzimer2005improving}
Alina Beygelzimer, Geoffrey Grinstein, Ralph Linsker, and Irina Rish.
\newblock Improving network robustness by edge modification.
\newblock \emph{Physica A: Statistical Mechanics and its Applications},
  357\penalty0 (3-4):\penalty0 593--612, 2005.

\bibitem[Fortuna et~al.(2011)Fortuna, Bonachela, and
  Levin]{fortuna2011evolution}
Miguel~A Fortuna, Juan~A Bonachela, and Simon~A Levin.
\newblock Evolution of a modular software network.
\newblock \emph{Proceedings of the National Academy of Sciences}, 108\penalty0
  (50):\penalty0 19985--19989, 2011.

\bibitem[Nguyen et~al.(2010)Nguyen, Adams, and Hassan]{nguyen2010studying}
Thanh~HD Nguyen, Bram Adams, and Ahmed~E Hassan.
\newblock Studying the impact of dependency network measures on software
  quality.
\newblock In \emph{2010 IEEE International Conference on Software Maintenance},
  pages 1--10. IEEE, 2010.

\bibitem[Decan et~al.(2018)Decan, Mens, and Constantinou]{decan2018impact}
Alexandre Decan, Tom Mens, and Eleni Constantinou.
\newblock On the impact of security vulnerabilities in the npm package
  dependency network.
\newblock In \emph{Proceedings of the 15th International Conference on Mining
  Software Repositories}, pages 181--191, 2018.

\bibitem[Guillaume et~al.(2004)Guillaume, Latapy, and
  Magnien]{guillaume2004comparison}
Jean-Loup Guillaume, Matthieu Latapy, and Cl{\'e}mence Magnien.
\newblock Comparison of failures and attacks on random and scale-free networks.
\newblock In \emph{International Conference on Principles of Distributed
  Systems}, pages 186--196. Springer, 2004.

\bibitem[Page et~al.(1999)Page, Brin, Motwani, and Winograd]{page1999pagerank}
Lawrence Page, Sergey Brin, Rajeev Motwani, and Terry Winograd.
\newblock The pagerank citation ranking: Bringing order to the web.
\newblock Technical report, Stanford InfoLab, 1999.

\bibitem[npm()]{npm_reg}
npm registry.
\newblock \url{https://docs.npmjs.com/cli/v7/using-npm/registry}.
\newblock Accessed: 2021-05-13.

\end{thebibliography}

\end{document}